# Real-Time Integrity Indices in Power Grid: A Synchronization Coefficient Based Clustering Approach

Hamzeh Davarikia
*Student Member, IEEE*
Louisiana State University
Baton Rouge, LA, USA

Masoud Barati
*Member, IEEE*
Louisiana State University
Baton Rouge, LA, USA

Faycal Znidi
*Student Member, IEEE*
University of Arkansas at Little Rock
Little Rock, AR, USA

Kamran Iqbal
*Senior Member, IEEE*
University of Arkansas at Little Rock
Little Rock, AR, USA

*Abstract* – We propose a new methodology based on modularity clustering of synchronization coefficient, to identify coherent groups of generators in the power grid in real-time. The method uses real-time integrity indices, i.e., the Generators Connectivity Index (GCI) that represents how generators are coherently strong within the groups, the Generator Splitting Index (GSI) that reveals to what extent the generators in different groups tend to swing against the other groups, and the System Separation Index (*SI*) which discloses the overall system separation status. We demonstrate how these integrity indices can be used to study the dynamic behavior of the power system. Furthermore, a comparison analysis is conducted between the synchronization coefficient (KS) and the generator rotor angle correlation coefficient (CC). The proposed indices demonstrate the dynamic behavior of power system following occurrence the faults and thus represent a promising approach in power system islanding studies. Our methodology is simple, fast, and computationally attractive. Simulation case performed on IEEE 118-bus systems demonstrates the efficacy of our approach.

*Index Terms*-- Generator coherency, integrity indices, modularity clustering, power system islanding, synchronization coefficient

## I. Introduction

Enormous development of energy demand has pushed bulk power systems to operate very close to their limits and hence made them more vulnerable to disturbances. Therefore, it has become crucial to develop effective analysis methods to guarantee the stability of the modern power systems.

Coherency analysis has been broadly used in stability studies, to lessen the computational effort by aggregating coherent generators into a unique equivalent area. In that instance of perturbation in multi-machine power systems, some of the machines exhibit similar responses to the disturbance which means generators that are closely coupled in an electrical sense tend to swing together in groups throughout disturbances. Specifically, they present same dynamic behavior in that their rotor angles and frequencies remain very similar throughout the time interval of interest. This means the angular difference between any two generators is approximately constant over a period. This characteristic, if used, can help perform online behavior analysis of the coherent groups of generators, which is a promising approach to the power system islanding detection and can be used as a predictor for system instability.

Traditionally, the number of coherent groups of generators is determined in advance. However, the formation of coherent groups of generators depends on the type and location of the disturbance. Thus, a power network may have the potential for forming different patterns of cohesive groups (i.e., one coherent group may detach into smaller groups or conversely multiple groups may join to build a bigger coherent group) following the occurrence of disturbances. Consequently, a pre-designed sequence of events for construction of coherent units may lead to an inadequate defensive strategy against forced blackouts. Moreover, coherency analysis traditionally is carried out offline. Nevertheless, with changes in the network configuration and system operating point, the groups of coherent generators tend to vary with time. Subsequently, involving a real-time operation to define the system coherency is a more robust method than identifying coherency based on offline studies.

Owing to the importance of identification of coherent areas for operation and control studies, numerous approaches have been introduced to tackle this challenging problem. In [1], a general wide-area coherency detection methodology based on Project Pursuit (PP) is proposed, and an index named as the projection cumulative contribution rate (PCCR) is presented to determine the dominant coherency modes and coherent groups under various disturbances. However, how to select an appropriate PCCR is still a problem to be further investigated.

In [2], a coherency identification method based on the agglomerative hierarchical clustering (HC) and the Wide Area Measurement Systems (WAMS) is proposed. However, this process could not identify the number of the coherent groups of generators correctly, especially when there is only one generator in a coherent group. In [3], a comparison analyses between the synchronization coefficient *KS* and the generators rotor angle correlation coefficient (*CC*) is done; the results show that using *KS* is more advantageous than using *CC*. Also, an online methodology for determining the coherent groups of generators, assisted by the modularity clustering is introduced. In [4], synchronization coefficient among generators, aided by a heuristic, is used to detect groups of coherent generators. Later, the strength of connection within and between groups is calculated. In [5], a coherency identification method based on linearization of non-equilibrium points is introduced, so that oscillation modes between generators can be obtained using the eigenvalues of the coefficient matrix of linearized models. In [6], a hierarchical clustering algorithm is used to the rate of change of the generator bus voltage phase angles, and the rate of the change of the means of the generator bus voltage phase angles to identify the coherent groups of generators. In [7], WAMS are implemented using Fast Fourier Transform (FFT) based spectral techniques. However, this method assumes that stationary and linear data is analyzed, an assumption not always justified.

The Partitioning Around Medoids (PAM) algorithm is applied to determine the optimal coherent groups that minimize the changes by dynamic reduction and is used in [8] to compare it with the algorithm proposed K-means in [9]. The K-means clustering algorithm is simple to implement and is fast and sensitive to changes. However, the K-means algorithm has some drawbacks such as selection of initial centroids or the number of groups and the number of iterations needed to find the clusters [10]. In [11], K-harmonic means clustering approach for online coherency analysis of synchronous generators has been presented. However, this method requires a second algorithm to determine the number of groups required during the online analysis; hence clusters may be merged or split to satisfy this threshold. In [13],[14], a new alternative and much more justified approach that uses *CC* as coherency index is presented. This tactic identifies coherent groups of generators based on the *CC* between rotor angle/speed oscillations of generators. This method is based on the heuristic methodology for partitioning the coherent groups in a predefined time window. However, one major drawback with heuristics is that we do not know if the result is close to optimal or not. Further, most of the aforementioned methods are computationally expensive and require upfront determination of the coherent groups, which is not always justifiable.

Traditionally, grouping the generators based on the calculated indices was done by the heuristics [4], [13-14]; however, there is no guarantee that the proposed heuristics work well in all situations. To address this problem, modularity clustering is introduced as an efficient grouping method. With this technique, not only the coherent groups of generators are determined online, but also the strength of coherency, network integrity, and the overall system separation status are analyzed. The method uses real-time integrity indices: the GCI index that represents how generators are coherently strong within the groups, the GSI index that reveals to what extent the generators in different groups tend to swing against the other groups, and the SI index which discloses the overall system separation status. Our methodology is computationally simple, fast, and it neither requires a predefined number of desired groups, nor a preset calculation time window. The major contributions of this work are as follow: 1- new real-time integrity indices in power system are proposed based on modularity clustering and synchronization coefficient. The introduced indices can be used for dynamic instability or uncontrolled islanding detection in bulk power networks. 2- A comprehensive numerical study is carried out to demonstrate the performance of the proposed indices in a large-scale test system.

The rest of this paper is structured as follows. Section II introduces a comparison analysis between the *KS* and the *CC*. Section III introduces the coherency problem and the basic concepts of modularity clustering algorithm. In section IV, the execution of the proposed method is discussed, and the online detection of coherent groups of generators and network integrity indices is determined. In section V, the proposed method is applied to the IEEE 118-bus test systems to demonstrate the validity of the proposed scheme.

## II. ONLINE GENERATOR COHERENCY DETERMINATION

Two efficient methods of generator coherency detection, namely correlation coefficient among generators rotor angles and synchronization coefficient among generators are investigated in this section.

### A. Correlation Characteristics of Generators Rotor Angle

Since the correlation is a measure of the strength of linear association between two variables, coherency among two different generators can be measured by the Pearson product-moment correlation coefficient of their rotor angle oscillations in a predefined time window [14]. Equation (1) represents the *CC* between two generators *i* and *j* based on their rotor angle oscillation in an arbitrary time window with *n* sample points:

$$CC_{ij} = \frac{n\sum \delta_i \delta_j - \sum \delta_i \sum \delta_j}{\sqrt{n\sum \delta_i^2 - \left(\sum \delta_i\right)^2}\sqrt{n\sum \delta_j^2 - \left(\sum \delta_j\right)^2}} \quad (1)$$

where $\delta_i = \{\delta_{i_1}, \delta_{i_2}, ..., \delta_{i_n}\}$ is the dataset of the rotor angle of the $i^{th}$ generator in the desired time window containing *n* values of rotor angle $\delta_i$. The *CC* is ranging from -1 to +1, in which -1 stands for the anti-phase rotor angles, whereas +1 represents the perfectly correlated generators. Sizing of the time window in the *CC* approach is one of its main drawback which makes this method less attractive in real-time applications.

### B. Synchronization Coefficient Among Generators

The *KS*, introduced in [4], is an appropriate index for evaluating generator coherency and generator dependency. Equation (2) shows the *KS* between generator *i* and *j*

$$KS_{ij} = \sum_{j=1}^{m} |E_i'| |E_j'| B_{ij} \cos(\delta_i - \delta_j) \quad (2)$$

where $|E_i'|$ is the per unit generator voltage behind the transient reactance, $B_{ij}$ represents the imaginary part of the reduced admittance matrix, and $\delta_i$ is the rotor angle of the $i^{th}$ machine. Since the *KS* is a function of generator voltage, admittance, and rotor angle, it does not have the *CC* drawbacks and not only returns the coherency index among the pairs of generators at any point in time, it measures the coherency among slack generator and other machines. Accordingly, in this paper, generator coherency identification is carried out using *KS* in the real-time studies. After computing the *KS* among all synchronous generators in the system at any point in time, there is a need to partition the generators into the coherent groups of generators. To address this problem, modularity clustering is introduced as an efficient grouping method in the next section.

### III. MODULARITY CLUSTERING ALGORITHM

To identify generators coherency, it is necessary to find strongly connected groups of generators, since the units that are strongly coupled tend to maintain synchronism. Online coherency detection based on modularity clustering algorithm will be used to achieve this purpose; it neither requires a predefined number of groups nor requires defining a threshold value. The objective of this method is to separate the network into groups of vertices that have weak connections between them and use the modularity clustering algorithm to measure the strength of division of the network. In this paper, the methodology proposed in [15], which has the $O(nlog2n)$ complexity, is used.

### IV. ONLINE DETECTION OF COHERENT GROUPS OF GENERATORS AND NETWORK INTEGRITY INDICES

The first step in evaluating generator coherency of a power network at any point in time is to calculate the *KS* among all the generators and form the Synchronization Coefficient Matrix (*KS* Matrix). Then, N groups of coherent generators can be achieved by applying modularity clustering on the *KS* Matrix. The power network integrity indices can be obtained by performing further analysis on the clustered *KS* Matrix; the $N \times N$ matrix of coherent groups (KSGM), which shows the overall coherency of the system, can then be extracted from the *KS* Matrix. The diagonal elements of the KSGM are the weighted average in each group in the *KS* Matrix, and the off-diagonal elements are the mutual weighted average between different groups of generators. The *GCI* is defined as the average of KSGM off-diagonal while *GSI* is the average of diagonal of KSGM. *GCI* represents how generators are coherently strong within the groups, while *GSI* shows to what extent the generators in different groups tend to swing against the other groups. The *SI* is also defined as *GCI* divided by *GSI*, which represents the overall system separation status.

### V. SIMULATION STUDIES

The IEEE 118-bus system [16] is employed to demonstrate the performance of the proposed approaches. To compare the performance of two methods, the associated *CC* and *KS* matrices are calculated at a specific time. Moreover, the proposed system evaluation indices are calculated, and their trend during the simulation studies will be explored. The IEEE 118-bus system with the load level of the 3668 MW, modeled in DIgSILENT PowerFactory, is shown in Fig. 1 [17]. Table I lists the events that occur in the test bed system during the simulation time of 30 sec.

TABLE I. EVENTS OCCURRED IN THE IEEE 118-BUS SYSTEM

| Time (sec) | Description |
|---|---|
| 0.01 | Load Event – Increasing the system load level by 1.5 times |
| 3.00 | Short circuit on lines 68-81, 30-38 , 45-49 , 46-47 , 46-48, 47-49 |
| 3.20 | Switch event on lines 68-81, 30-38 , 45-49 , 46-47 , 46-48, 47-49 |

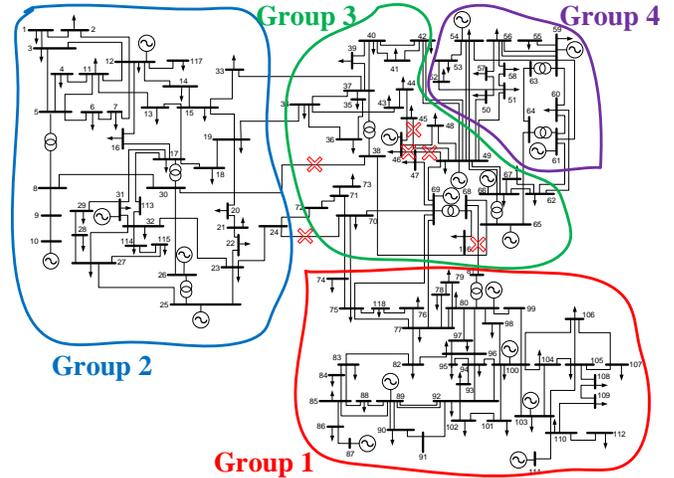

Figure 1. IEEE 118-bus system. The crosses show the events

Fig. 2 shows the generator rotor angle oscillation during the simulation study and in a specific time window between $5.5 \leq t \leq 9.5$ sec. Obviously, the rotor angle oscillations are damped, and all of the machines remain in synchronism while groups of generators became stronger following the events. As can be seen in Fig. 2(a) the power system remains stable following occurrence the disturbances, and hence the grid integrity indices *GCI*, *GSI*, and *SI* reach their steady points after $t = 23$ sec. Calculating the CC and KS for all pairs of generators result in the KS Matrix and CC Matrix shown in Fig.7 and Fig.8 respectively. Applying Modularity Clustering on the *KS* Matrix shown in Fig. 7, results in four separate coherent groups of generators: 1: {G14,G15,G16,G17,G18, G19}, 2: {G1,G2,G3,G4,G5}, 3: {G6,G7,G11,G12,G13} , 4: {G8,G9,G10}, and On the other hand, the resulted groups of the *CC* Matrix are only two groups as depicted in Fig. 8.

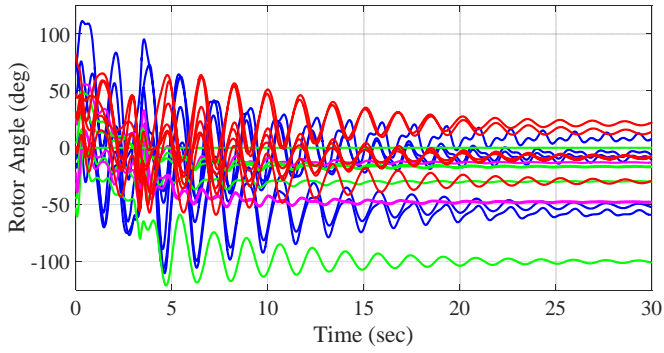

(a)

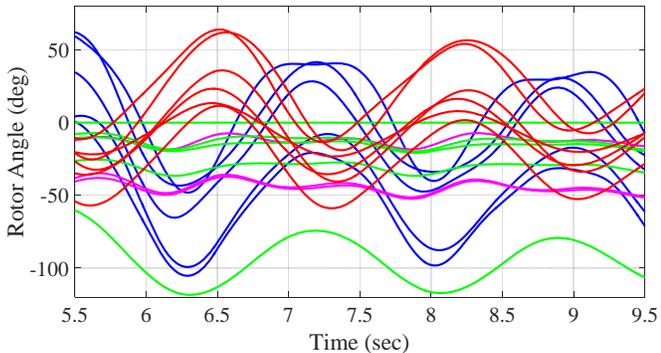

(b)

Figure 2.  (a) Rotor angles oscillation during study period, (b) Rotor angle oscillation during the specific time window

|  | **Group1** | **Group2** | **Group3** | **Group4** |
|---|---|---|---|---|
| **Group1** | 4.371 | 0.008 | 0.388 | 0.000 |
| **Group2** | 0.008 | 6.124 | 0.202 | 0.000 |
| **Group3** | 0.388 | 0.202 | 8.521 | 3.133 |
| **Group4** | 0.000 | 0.000 | 3.133 | 14.105 |

Figure 3.  KSGM corresponded to KS Matrix at t = 7.25 sec

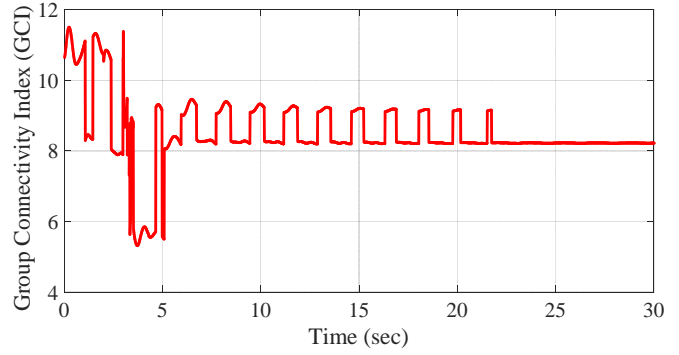

Figure 4.  Group Connectivity Index during study period

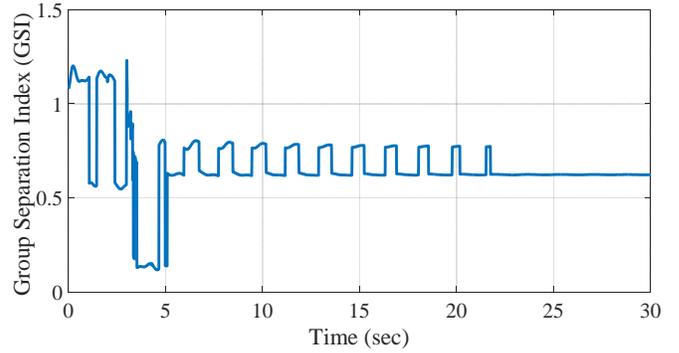

Figure 5.  Group Separation Index during study period

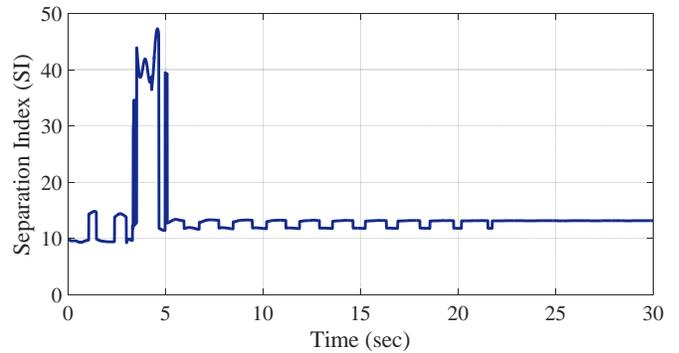

Figure 6.  Separation Index during study period

|  | G1 | G2 | G3 | G4 | G5 | G6 | G7 | G11 | G12 | G13 | G8 | G9 | G10 | G14 | G15 | G16 | G17 | G18 | G19 |
|---|---|---|---|---|---|---|---|---|---|---|---|---|---|---|---|---|---|---|---|
| **G1** |  | 7.67 | 0.32 | 2.57 | 0.88 | 0.00 | 0.07 | 0.16 | 0.00 | 0.01 | 0.00 | 0.00 | 0.00 | 0.00 | 0.00 | 0.00 | 0.00 | 0.00 | 0.00 |
| **G2** | 7.67 |  | 1.01 | 3.49 | 3.55 | 0.20 | 0.56 | 1.34 | 0.00 | 0.09 | 0.00 | 0.00 | 0.00 | 0.02 | 0.00 | 0.00 | 0.00 | 0.00 | 0.00 |
| **G3** | 0.32 | 1.01 |  | 34.24 | 6.08 | 0.00 | 0.08 | 0.20 | 0.00 | 0.79 | 0.00 | 0.00 | 0.00 | 0.13 | 0.00 | 0.01 | 0.00 | 0.00 | 0.00 |
| **G4** | 2.57 | 3.49 | 34.24 |  | 1.42 | 0.00 | 0.10 | 0.26 | 0.00 | 0.02 | 0.00 | 0.00 | 0.00 | 0.00 | 0.00 | 0.00 | 0.00 | 0.00 | 0.00 |
| **G5** | 0.88 | 3.55 | 6.08 | 1.42 |  | 0.07 | 0.23 | 0.57 | 0.00 | 0.30 | 0.00 | 0.00 | 0.00 | 0.06 | 0.00 | 0.00 | 0.00 | 0.00 | 0.00 |
| **G6** | 0.00 | 0.20 | 0.00 | 0.00 | 0.07 |  | 0.29 | 0.63 | 0.00 | 0.00 | 0.00 | 0.00 | 0.00 | 0.00 | 0.00 | 0.00 | 0.00 | 0.00 | 0.00 |
| **G7** | 0.07 | 0.56 | 0.08 | 0.10 | 0.23 | 0.29 |  | 1.65 | 26.82 | 3.64 | 15.37 | 0.67 | 0.00 | 0.00 | 0.00 | 0.00 | 0.00 | 0.00 | 0.00 |
| **G11** | 0.16 | 1.34 | 0.20 | 0.26 | 0.57 | 0.63 | 1.65 |  | 30.55 | 21.63 | 0.00 | 8.28 | 15.03 | 0.00 | 0.00 | 0.00 | 0.00 | 0.00 | 0.00 |
| **G12** | 0.00 | 0.00 | 0.00 | 0.00 | 0.00 | 0.00 | 26.82 | 30.55 |  | 0.00 | 0.00 | 0.27 | 7.37 | 0.00 | 0.00 | 0.00 | 0.00 | 0.00 | 0.00 |
| **G13** | 0.01 | 0.09 | 0.79 | 0.02 | 0.30 | 0.00 | 3.64 | 21.63 | 0.00 |  | 0.00 | 0.00 | 0.00 | 10.56 | 0.08 | 0.57 | 0.40 | 0.00 | 0.00 |
| **G8** | 0.00 | 0.00 | 0.00 | 0.00 | 0.00 | 0.00 | 15.37 | 0.00 | 0.00 | 0.00 |  | 18.64 | 0.00 | 0.00 | 0.00 | 0.00 | 0.00 | 0.00 | 0.00 |
| **G9** | 0.00 | 0.00 | 0.00 | 0.00 | 0.00 | 0.00 | 0.67 | 8.28 | 0.27 | 0.00 | 18.64 |  | 23.68 | 0.00 | 0.00 | 0.00 | 0.00 | 0.00 | 0.00 |
| **G10** | 0.00 | 0.00 | 0.00 | 0.00 | 0.00 | 0.00 | 0.00 | 15.03 | 7.37 | 0.00 | 0.00 | 23.68 |  | 0.00 | 0.00 | 0.00 | 0.00 | 0.00 | 0.00 |
| **G14** | 0.00 | 0.02 | 0.13 | 0.00 | 0.06 | 0.00 | 0.00 | 0.00 | 0.00 | 10.56 | 0.00 | 0.00 | 0.00 |  | 0.42 | 3.86 | 13.89 | 0.00 | 0.00 |
| **G15** | 0.00 | 0.00 | 0.00 | 0.00 | 0.00 | 0.00 | 0.00 | 0.00 | 0.00 | 0.08 | 0.00 | 0.00 | 0.00 | 0.42 |  | 1.56 | 0.14 | 0.00 | 0.00 |
| **G16** | 0.00 | 0.00 | 0.01 | 0.00 | 0.00 | 0.00 | 0.00 | 0.00 | 0.00 | 0.57 | 0.00 | 0.00 | 0.00 | 3.86 | 1.56 |  | 10.32 | 0.00 | 0.00 |
| **G17** | 0.00 | 0.00 | 0.00 | 0.00 | 0.00 | 0.00 | 0.00 | 0.00 | 0.00 | 0.40 | 0.00 | 0.00 | 0.00 | 13.89 | 0.14 | 10.32 |  | 29.74 | 0.95 |
| **G18** | 0.00 | 0.00 | 0.00 | 0.00 | 0.00 | 0.00 | 0.00 | 0.00 | 0.00 | 0.00 | 0.00 | 0.00 | 0.00 | 0.00 | 0.00 | 0.00 | 29.74 |  | 4.67 |
| **G19** | 0.00 | 0.00 | 0.00 | 0.00 | 0.00 | 0.00 | 0.00 | 0.00 | 0.00 | 0.00 | 0.00 | 0.00 | 0.00 | 0.00 | 0.00 | 0.00 | 0.95 | 4.67 |  |

Figure 7.  KS Matrix in t = 7.25 sec

|     | G1 | G2 | G3 | G4 | G5 | G6 | G7 | G11 | G12 | G8 | G9 | G10 | G14 | G15 | G16 | G17 | G18 | G19 |
|-----|----|----|----|----|----|----|----|-----|-----|----|----|-----|-----|-----|-----|-----|-----|-----|
| G1  |      | 0.99 | 0.93 | 0.94 | 0.97 | 0.98 | 0.99 | 0.96 | 0.95 | 0.98 | 0.99 | 0.98 | 0.00 | 0.00 | 0.00 | 0.00 | 0.00 | 0.00 |
| G2  | 0.99 |      | 0.96 | 0.96 | 0.98 | 0.99 | 0.96 | 0.93 | 0.92 | 0.96 | 0.97 | 0.96 | 0.00 | 0.00 | 0.00 | 0.00 | 0.00 | 0.00 |
| G3  | 0.93 | 0.96 |      | 0.99 | 0.95 | 0.99 | 0.88 | 0.84 | 0.83 | 0.89 | 0.89 | 0.88 | 0.00 | 0.00 | 0.00 | 0.00 | 0.00 | 0.00 |
| G4  | 0.94 | 0.96 | 0.99 |      | 0.96 | 0.99 | 0.90 | 0.87 | 0.86 | 0.92 | 0.91 | 0.91 | 0.00 | 0.00 | 0.00 | 0.00 | 0.00 | 0.00 |
| G5  | 0.97 | 0.98 | 0.95 | 0.96 |      | 0.98 | 0.97 | 0.96 | 0.96 | 0.98 | 0.97 | 0.97 | 0.00 | 0.00 | 0.00 | 0.00 | 0.00 | 0.00 |
| G6  | 0.98 | 0.99 | 0.99 | 0.99 | 0.98 |      | 0.94 | 0.91 | 0.90 | 0.95 | 0.95 | 0.95 | 0.00 | 0.00 | 0.00 | 0.00 | 0.00 | 0.00 |
| G7  | 0.99 | 0.96 | 0.88 | 0.90 | 0.97 | 0.94 |      | 0.99 | 0.99 | 1.00 | 1.00 | 1.00 | 0.00 | 0.00 | 0.00 | 0.00 | 0.00 | 0.00 |
| G11 | 0.96 | 0.93 | 0.84 | 0.87 | 0.96 | 0.91 | 0.99 |      | 1.00 | 0.99 | 0.99 | 0.99 | 0.00 | 0.00 | 0.00 | 0.00 | 0.00 | 0.00 |
| G12 | 0.95 | 0.92 | 0.83 | 0.86 | 0.96 | 0.90 | 0.99 | 1.00 |      | 0.99 | 0.98 | 0.99 | 0.00 | 0.00 | 0.00 | 0.00 | 0.00 | 0.00 |
| G8  | 0.98 | 0.96 | 0.89 | 0.92 | 0.98 | 0.95 | 1.00 | 0.99 | 0.99 |      | 0.99 | 1.00 | 0.00 | 0.00 | 0.00 | 0.00 | 0.00 | 0.00 |
| G9  | 0.99 | 0.97 | 0.89 | 0.91 | 0.97 | 0.95 | 1.00 | 0.99 | 0.98 | 0.99 |      | 1.00 | 0.00 | 0.00 | 0.00 | 0.00 | 0.00 | 0.00 |
| G10 | 0.98 | 0.96 | 0.88 | 0.91 | 0.97 | 0.95 | 1.00 | 0.99 | 0.99 | 1.00 | 1.00 |      | 0.00 | 0.00 | 0.00 | 0.00 | 0.00 | 0.00 |
| G14 | 0.00 | 0.00 | 0.00 | 0.00 | 0.00 | 0.00 | 0.00 | 0.00 | 0.00 | 0.00 | 0.00 | 0.00 |      | 0.99 | 0.94 | 0.96 | 1.00 | 0.99 |
| G15 | 0.00 | 0.00 | 0.00 | 0.00 | 0.00 | 0.00 | 0.00 | 0.00 | 0.00 | 0.00 | 0.00 | 0.00 | 0.99 |      | 0.92 | 0.96 | 0.99 | 0.98 |
| G16 | 0.00 | 0.00 | 0.00 | 0.00 | 0.00 | 0.00 | 0.00 | 0.00 | 0.00 | 0.00 | 0.00 | 0.00 | 0.94 | 0.92 |      | 0.99 | 0.97 | 0.98 |
| G17 | 0.00 | 0.00 | 0.00 | 0.00 | 0.00 | 0.00 | 0.00 | 0.00 | 0.00 | 0.00 | 0.00 | 0.00 | 0.96 | 0.96 | 0.99 |      | 0.98 | 0.99 |
| G18 | 0.00 | 0.00 | 0.00 | 0.00 | 0.00 | 0.00 | 0.00 | 0.00 | 0.00 | 0.00 | 0.00 | 0.00 | 1.00 | 0.99 | 0.97 | 0.98 |      | 1.00 |
| G19 | 0.00 | 0.00 | 0.00 | 0.00 | 0.00 | 0.00 | 0.00 | 0.00 | 0.00 | 0.00 | 0.00 | 0.00 | 0.99 | 0.98 | 0.98 | 0.99 | 1.00 |      |

Figure 8. CC Matrix in t = 7.25 sec, with n =100

This fact can be traced in the KSGM associated to *KS* Matrix demonstrated in Fig. 3, in which the coherency between the groups 3 and 4 is large when compared with the coherency between group 1 and groups 2, 3 and 4. It means one can consider groups 3 and 4 (and 2) as one group if the resolution of the grouping is not fine enough. Fig. 2(b) can prove the accuracy of the *KS* Matrix grouping results. The trend of online integrity indices shown in Fig. 4, Fig. 5, and Fig. 6 clearly shows the dynamical behavior of groups of generators. Since the faults are not damaging in our case study, the indices, which can be used to predict system power system islanding, will reach to the steady point after some oscillations.

## VI. CONCLUSION

A comparative investigation between the synchronizing coefficient matrix (*KS*) and the generator rotor angle correlation coefficients (*CC*) was accomplished in this paper. The results show that using *KS* is more advantageous than using *CC* in the real-time identification of coherent groups. In addition, an online methodology for determining the coherent groups of generators assisted by the modularity clustering was introduced. This approach not only accomplishes real-time coherency detection without a predefined number of coherent groups or specifying a time-window for study, but also provides the online indices for tracking the dynamic behavior of the modern power system. The proposed indices can also be used in predicting of uncontrolled islanding.

REFERENCES

[1] T. Jiang, H, J. Jia, H. Y. Yuan, et al, "Projection pursuit: A general methodology of wide-area coherency detection in bulk power grid," IEEE Transactions on Power Systems, vol. 31, no. 4, pp. 2776-2786, 2016
[2] Z. Z. Lin, F. S. Wen, J. H. Zhao, et al, "Controlled islanding schemes for interconnected power systems based on coherent generator group identification and wide-area measurements," Journal of Modern Power Systems and Clean Energy, vol. 4, no. 3, pp. 440-453, 2016
[3] F. Znidi, H. Davarikia and K. Iqbal, "Modularity clustering based detection of coherent groups of generators with generator integrity indices," in 2017 IEEE Power & Energy Society General Meeting, 2017, pp. 1-5.
[4] H. Davarikia, F. Znidi, M. R. Aghamohammadi, K. Iqbal. Identification of coherent groups of generators based on synchronization coefficient. 2016 IEEE Power and Energy Society General Meeting (PESGM). 2016:1-5.
[5] Ma, Zhenbin, Lei Ding, Zhifan Liu, Yichen Guo, Qian Liu, and Weiyu Bao. "The Application of a Generator Coherency Identification Method Based on Linearization in Complex Power System." Presented at the 2016 China International Conference on Electricity Distribution (CICED), August 10-13, 2016.
[6] Ali, M., Mork, B.A., Bohmann, L.J., Brown, L.E.: "Detection of coherent groups of generators and the need for system separation using synchrophasor data". Proc. Int. Conf. on Power Engineering and Optimization, June 2013, pp. 7–12.
[7] M. Jonsson, M. Begovic, and J. Daalder, "A new method suitable for real-time generator coherency determination," IEEE Trans. Power Syst., vol. 19, no. 3, pp. 1473–1482, Aug. 2004.
[8] Pyo, G.C., Park, J.W., and Moon, S.I. "A new method for dynamic reduction of power system using PAM algorithm," in Proc. Power and Energy Society General Meeting, 2010 IEEE, July 2010.
[9] S. K. Joo, C. C. Liu, L. E. Jones, and J. W. Choe "Coherency and aggregation techniques incorporating rotor and voltage dynamics" IEEE Trans. on Power Systems., vol. 19, no. 2, pp.1068-1075, May 2004.
[10] Kalpana D. Joshi et al., "Modified K-Means for Better Initial Cluster Centres", International Journal of Computer Science and Mobile Computing , IJCSMC, Vol. 2, Issue. 7, July 2013, pg.219– 223.
[11] J. A. Taylor and S. M. Halpin, "Approximation of Generator Coherency Based on Synchronization Coefficients," in System Theory, 2007. SSST '07. Thirty-Ninth Southeastern Symposium on, 2007, pp. 47-51.
[12] Farkhondeh Jabari , Heresh Seyedi , Sajad Najafi Ravadanegh "Online Aggregation of Coherent Generators Based on Electrical Parameters of Synchronous Generators". International Journal of Smart Electrical Engineering, winter 2015.
[13] M.R. Aghamohammadi, S.M. Tabandeh, "A new approach for online coherency identification in power systems based on correlation characteristics of generators rotor oscillations". Electrical Power and Energy Systems, 2016.
[14] Aghamohammadi, M.R., Tabandeh, S.M.: "Online coherency identification based on correlation characteristics of generator rotor angles". Proc. Int. Conf. on Power and Energy, 2012, pp. 499–504.
[15] Fortunato, Santo. "Community detection in graphs." Physics reports 486.3-5 (2010): 75-174..
[16] P. Demetriou, "Dynamic IEEE test systems for transient analysis", IEEE Syst. J., vol. PP, no. 99, pp. 1-7, Jul. 2015.
[17] H. Davarikia, "Investment plan against malicious attacks on power networks: multilevel game-theoretic models with shared cognition," MSc. dissertation, Dept. Systems Engineering., Univ. Arkansas at Little Rock, Little Rock, 2017.